\documentclass[aip,graphicx]{revtex4-1}
\usepackage{geometry}
\usepackage{setspace}
\usepackage{fullpage}
\usepackage{color}
\usepackage{hyperref}
\usepackage[pdftex]{graphicx}
\draft 

\begin{document}


\title{Grain boundary characteristics of Fe-based superconductors} 

\author{Kazumasa\,Iida}\email{iida@mp.pse.nagoya-u.ac.jp}
\affiliation{Department of Materials Physics, Nagoya University, Furo-cho, Chikusa-ku, Nagoya 464-8603, Japan}
\affiliation{JST CREST, Sanbancho 5, Chiyoda-ku, Tokyo 102-0075, Japan}
\author{Jens\,H\"{a}nisch}
\affiliation{Institute for Technical Physics, Karlsruhe Institute of Technology, Hermann-von-Helmholtz-Platz 1, 76344 Eggenstein-Leopoldshafen, Germany}
\author{Akiyasu\,Yamamoto}
\affiliation{Department of Applied Physics, Tokyo University of Agriculture and Technology, 2-24-16 Naka-cho, Koganei-shi, Tokyo 184-8588, Japan}
\affiliation{JST CREST, Sanbancho 5, Chiyoda-ku, Tokyo 102-0075, Japan}


\date{\today}

\begin{abstract}
Understanding the nature of grain boundary (GB) characteristics in combination with creating low-energy GBs by modifying the processing conditions, so-called GB engineering, is of great importance for controlling and reducing the defect density, leading to improved functionalities of polycrystalline metals and ceramics. For superconductors particularly, including both low- and high-temperature superconductors, GB engineering has been developed to improve especially the critical current densities, $J_{\rm c}$, across these GBs. The intrinsic physical properties of a given superconductor such as the coherence length, the order parameter symmetry, and their anisotropies would determine the strategy of GB engineering. In this topical review, we present an overview of the GB characteristics and GB engineering of Fe-based superconductors (FBS) in the form of polycrystalline bulks and wires, and thin films with application potential, e.g. for high-field magnet wires. Prior to the FBS, GB engineering of the cuprates and MgB$_2$ are also briefly covered.
\end{abstract}

\maketitle
\section{Introduction}
Boundaries between adjacent crystallites of typically differing orientation and/or composition are called $grain$ $boundaries$, where the atomic arrangement is in disorder. A grain boundary (GB) with a high interfacial energy (i.e. usually two adjacent grains having a large misorientation angle) is often the origin of macroscopic defects such as cracks and erosion damage, and these defects propagate along the GBs. On the other hand, GBs with a low interfacial energy [e.g. low-angle GBs and coincidence boundaries like twin boundaries] are hardly the origin of macroscopic defects. In low-angle GBs, dense dislocation arrays are formed with a spacing $D$ given by Frank$^\prime$s formula:
 
\begin{equation}
D = \left(\frac{b}{2}\right) / {\rm sin}\left(\frac{\theta_{\rm GB}}{2}\right)\sim \frac{b}{\theta_{\rm GB}}
\end{equation}

\noindent
where $b=|\vec{b}|$ is the norm of the Burgers vector (for symmetric [001]-tilt GBs in bicrystal films of e.g. cuprates and pnictides as discussed in this review,
$|\vec{b}|$ is the in-plane lattice parameter) and $\theta_{\rm GB}$ is the misorientation angle between the two adjacent grains.

For both low- and high-temperature superconductors (LTS and HTS), GBs affect their properties: For instance, macroscopic defects (cracks) impede the super-current flow. In this case, GBs are detrimental defects to the critical current density $J_{\rm c}$ and, hence, should be avoided.

For cuprates, not only macroscopic but also microscopic defects are problematic due to the short coherence length $\xi$ (less than 2\,nm for $ab$-plane)\,\cite{Tomimoto} and the anisotropic superconducting order parameter ($d$-wave)\,\cite{Tsuei}. If the strained regions around the dislocations within the GB touch each other, the low-angle GB turns into a high-angle GB. This angle is around $10^\circ \sim 15^\circ$. However, in cuprates a wider region around the centre of the dislocation is perturbed electronically and the inter-grain $J_{\rm c}$ starts to decrease exponentially already for $\theta_{\rm GB}<10^\circ$. Also when the distance between strained regions is smaller or only slightly larger than the value of $\xi$ within the GB plane, the superconducting order parameter is depressed at the GB and, hence, the super-current flow is impeded. Additionally, bending of the electronic bands at GBs deplete the charge carrier density\,\cite{Hilgenkamp-1}, which seriously decreases the superconducting transition temperature $T_{\rm c}$. Hence, to achieve large $J_{\rm c}$ for cuprates, the crystallites should not only be aligned out-of-plane because $J_{\rm c}$ is anisotropic but also be aligned in-plane within the misorientation angle $\theta_{\rm GB}$, where the distance between strained regions is larger than $\xi_{ab}$ ($\xi_{ab}$: in-plane coherence length) is satisfied. On the other hand, for LTS and MgB$_2$ with large coherence length (e.g. 5$\sim$6\,nm) and $s$-wave symmetry (for MgB$_2$ anisotropic $s$-wave), GBs are able to pin vortices in the mixed state, and especially also GBs with rather large angles. Therefore, increasing the GB densities by reducing the grain size usually leads to improved in-field $J_{\rm c}$ properties.

As stated above, the symmetry of superconducting order parameter and the coherence length mainly govern the $J_{\rm c}$ characteristics at GBs for a given superconductor. Hence, for both LTS and HTS, understanding the nature of GB characteristics in combination with modifying the processing conditions, so-called GB engineering, is of great importance for reducing and/or controlling the defect density to tune the superconducting properties.

The Fe-based superconductors (FBS) have similar physical properties as the cuprates, for instance a short coherence length due to the small carrier concentration and a low Fermi velocity\,\cite{Putti}. Hence, FBS also seem to share the same GB issues as the cuprates. However, $J_{\rm c}$ across GBs in FBS is not as severely reduced due to several features, which is reviewed in this article. Many reviews of the synthesis of FBS in the form of bulk samples and thin films, and of their physical properties have been published to date. Additionally, excellent review articles concerning bicrystals and GBs in cuprates have already been published\,\cite{Hilgenkamp-1, Chan, John-1}, however, only a few on GB issues for FBS have been published (e.g.\,\onlinecite{John-2, Jens-Kazu}).

In the following sections, we present an overview of the GB characteristics and GB engineering of the FBS in the form of polycrystalline bulks and wires as well as thin films with application potential, e.g. for high-field magnet wires and compact bulks. In section I\hspace{-.1em}I, GB engineering of cuprates will be briefly summarised on the basis of the GB characteristics. Section I\hspace{-.1em}I\hspace{-.1em}I will describe how GBs affect the superconducting properties of MgB$_2$. These sections I\hspace{-.1em}I and I\hspace{-.1em}I\hspace{-.1em}I are relevant for understanding GB issues of FBS. Section I\hspace{-.1em}V will present the experimental reports on the misorientation angle dependence of the inter-grain $J_{\rm c}$ for various FBS films grown on symmetric [001]-tilt bicrystal substrates. In section V, GB engineering of FBS will be presented in the forms of both thin films and polycrystalline samples. The former part will describe how small misorientation angles affect the transport $J_{\rm c}$ of various FBS films deposited on technical substrates used for the second-generation coated conductors. The latter will present how to improve the superconducting properties of polycrystalline bulk and wire samples. Finally, we will discuss which kind of experiments are desired for better understanding the GB characteristics and summarise the review.

\section{Grain boundary engineering of cuprates}

\begin{figure}[t]
	\centering
		\includegraphics[width=7cm]{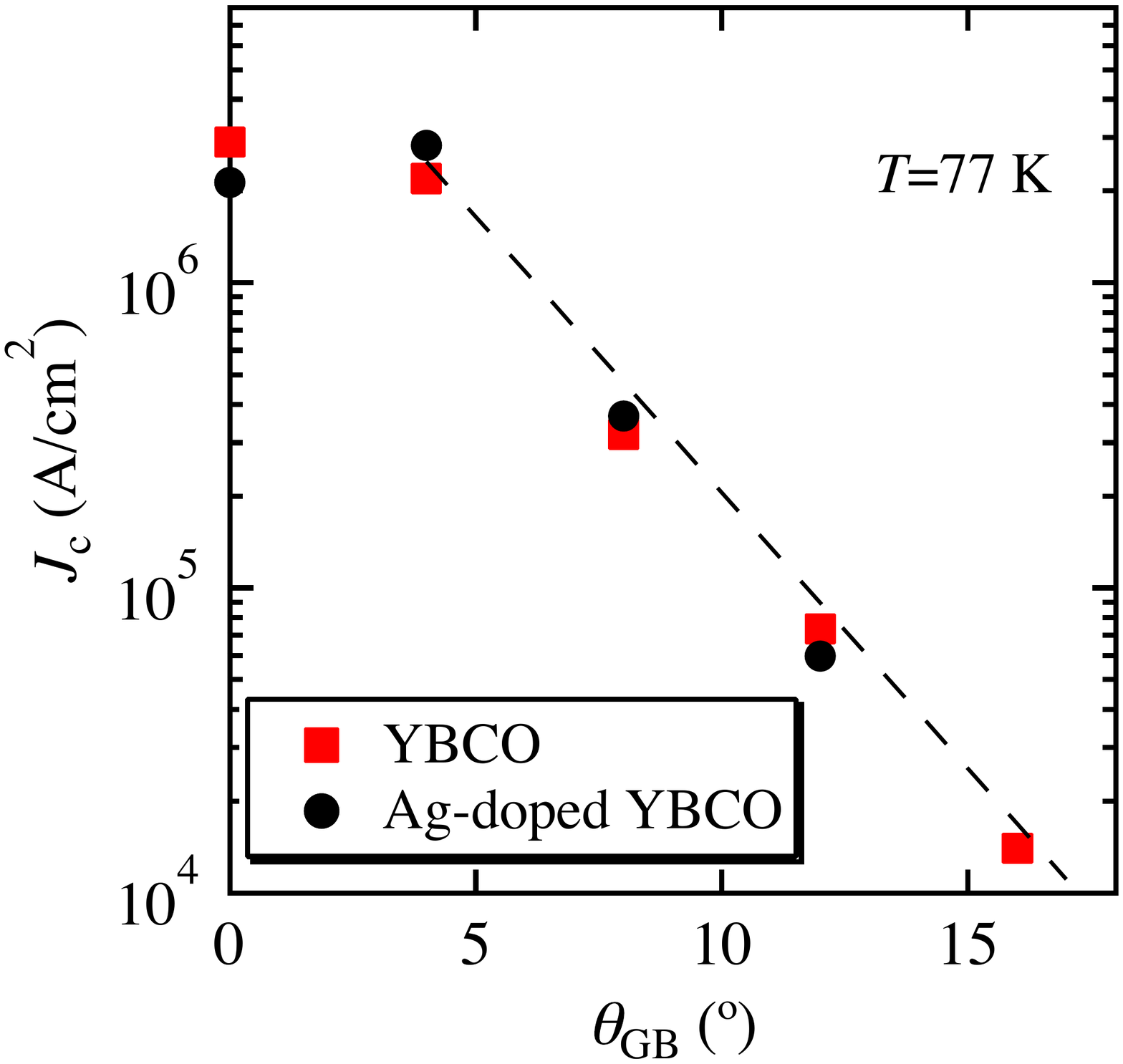}
		\caption{Critical current density at 77\,K of undoped and Ag-doped YBa$_2$Cu$_3$O$_7$ (YBCO) films grown on 
		SrTiO$_3$ [001]-tilt bicrystals of different misorientations. The data are taken from \,\onlinecite{Holzapfel}. The dashed line is a fit by eq. (2).} 
\label{fig:figure1}
\end{figure}

Thanks to $T_{\rm c}$ values higher than the boiling point of liquid nitrogen, the discovery of cuprates gave a significant excitement on our community. However, some material scientists were pessimistic about cuprates, since polycrystalline $RE$Ba$_2$Cu$_3$O$_7$ ($RE$: rare earth elements, $RE$BCO) was able to carry only $\sim$100\,A/cm$^2$ even at 4.2\,K\,\cite{Seuntjens}, which is far below the level needed for applications. A fundamental problem is that GBs with misorientation angle greater than 3$^\circ$ (this angle is called the critical angle, $\theta_{\rm c}$) constitute weak links. Figure\,\ref{fig:figure1} exemplifies the misorientation angle dependence of $J_{\rm c}$ for YBCO and Ag-doped YBCO grown on [001]-tilt symmetric SrTiO$_3$ bicrystal substrates\,\cite{Holzapfel}. Above $\theta_{\rm c}$, the angle dependence of inter-grain $J_{\rm c}$ ($J_{\rm c}^{\rm GB}$) is empirically described by

\begin{equation}
J_{\rm c}^{\rm GB}(\theta_{\rm GB}) =  J_{\rm c}^{\rm G}{\rm exp}\left(-\frac{\theta_{\rm GB}-\theta_{\rm c}}{\theta_{\rm 0}}\right)
\end{equation}

\noindent
where $J_{\rm c}^{\rm G}$ is the intra-grain critical current density and $\theta_{\rm 0}$ is the characteristic angle\,\cite{Holzapfel}. The data were well described by the equation above with $\theta_{\rm c}$=4$^\circ$ and $\theta_{\rm 0}$=2.4$^\circ$. Also for other cuprates (e.g. Nd$_{1.85}$Ce$_{0.15}$CuO$_{4-y}$\,\cite{Schoop} and Bi$_2$Sr$_2$Ca$_2$Cu$_3$O$_{10+\delta}$\,\cite{Jens-1}) inter-grain $J_{\rm c}$ decreases exponentially with misorientation angle.

Local strain\,\cite{Gurevich-1, Song, Deutscher} and bending of the electronic band structure\,\cite{Mannhart} resulting from the low carrier density combined with the large dielectric constant reduce the charge carrier density around GBs. As a result, dislocations or whole GB planes turn into antiferromagnetic insulators (i.e. Mott insulator). 

Several models for the microscopic understanding of the transport properties across the GBs for YBCO have been proposed to date. Gurevich and Pashitskii concluded that the misorientation-angle dependence of inter-grain $J_{\rm c}$ is mainly determined by the progressive decrease in the cross-sectional area of superconductivity, which is induced by local strain\,\cite{Gurevich-1}. They calculated the critical angle $\theta_{\rm c}$ to be around 3$^\circ-5^\circ$ at 77\,K, which is in good agreement with the experimental data. The progressive decrease in superconductivity around GBs is due to the shift of the Fermi level. The Fermi level of the GB regions having a thickness of the order of the Thomas-Fermi screening length lifts up around 5--10\,meV due to an excess of ion charge. Such small shifts are strong enough for a phase transition from superconductivity to antiferromagnetism to occur. This model explained the misorientation-angle dependence of inter-grain $J_{\rm c}$ for the low-angle regime. However, the regions of enhanced strain and possible off-stoichiometries of adjacent dislocations touch and overlap for misorientation angles larger than $\sim$10$^\circ$ and, hence, the model cannot be applied for this angular regime. It is worth mentioning that the symmetry of the superconducting order parameter (isotropic $s$- and anisotropic $d$-wave) seems not to affect the angle dependence of $J_{\rm c}$ near the critical angle.

Another calculation based on the bond contraction model showed that the pairing is weakened by the tensile strain surrounding edge dislocations\,\cite{Deutscher}. This explains the exponential decay of inter-grain $J_{\rm c}$. Kasatkin $et$ $al$. calculated the inter-grain $J_{\rm c}$ for [001]-tilt GBs on the assumption of the periodic pinning potential along the GBs\,\cite{Kasatkin}. They calculated the inter-grain $J_{\rm c}$ as a function of the in-plane coherence length and the norm of the Burgers vector, $b$, (as stated before, this value corresponds to the in-plane lattice parameter). In this model, the critical angle increases for decreasing ratio of $\xi_{ab}/b$. The results agree well with the experimental data for YBCO bicrystals with low misorientation angles. However, as pointed out by the authors, this model cannot explain the larger critical angle of FBS.

For arbitrary misorientation-angles, the inter-grain $J_{\rm c}$ has been calculated by Graser $et$ $al$\,\cite{Graser}. They initially simulated the atomic arrangement of YBCO at grain boundaries using a molecular dynamics procedure. Then a tight-binding model including the $d_{x^2-y^2}$ wave symmetry with parameters based on the atomic arrangement determined before was constructed for calculation. This calculation showed also good agreement with the experimental results.

As stated above, GB regions turn into insulating layers, which cannot be explained by the $d_{x^2-y^2}$ wave symmetry of the order parameter alone as pointed out by Schmehl $et$ $al$\,\cite{Schmehl}. The product of the grain-boundary normal state resistance $R_{\rm n}$ and the grain boundary area $A$ is in the range 10$^{-9}$$\leq$$R_{\rm n}A$$\leq$10$^{-8}$ $\Omega$cm$^2$ at 4.2\,K for cuprates\,\cite{Hilgenkamp-2}, which is one or two orders of magnitude higher than for Fe-based superconductors. To overcome these problems, Schmehl $et$ $al$. tried to increase the charge carrier density by partially substituting Y$^{+3}$ with Ca$^{+2}$ in YBCO\,\cite{Schmehl}. As a result, the inter-grain $J_{\rm c}$ for [001]-tilt Y$_{0.7}$Ca$_{0.3}$Ba$_2$Cu$_3$O$_{7-\delta}$ bicrystal films with $\theta_{\rm GB}$=24$^\circ$ was enhanced by a factor of around 8 at 4.2\,K compared to non-doped YBCO. Indeed, $E$($J$) curves at the same reduced temperature ($t$=$T/T_{\rm c}$=0.85) for Ca-doped YBCO fabricated on [001]-tilt bicrystal substrate with $\theta_{\rm GB}$=5$^\circ$ did not show the non-ohmic linear differential behavior (i.e. the region where electric field $E$ increases linearly with the current density $J$) known from pure YBCO\,\cite{Daniels}. Non-ohmic linear differential behavior is caused by viscous flux flow along the grain boundaries\,\cite{Diaz} and hence a clear indication that $J_{\rm c}$ is limited by GBs. Although Ca-doping mitigates the reduction of inter-grain $J_{\rm c}$, the superconducting transition temperature is decreased to the range of 68$\leq$$T_{\rm c}$$\leq$79\,K, when Ca is doped not only in the GB region but also in the adjacent grains. In order to dope Ca only locally whilst grains are kept high-$T_{\rm c}$, Hammerl $et$ $al$. fabricated a heterostructure of 
YBCO/Y$_{1-x}$Ca$_{x}$Ba$_2$Cu$_3$O$_{7-\delta}$ on 24$^\circ$ [001]-tilt SrTiO$_3$ bicrystal\,\cite{Hammerl}. As a result, Ca preferentially diffused  in the GB region and thereby doped the sample locally at the GB. The resultant films showed a self-field $J_{\rm c}$ as high as 3.3$\times$10$^5$\,A/cm$^2$ at 77\,K, which is almost the same level as for 7$^\circ$ bicrystal films.

Another possible method of increasing the carrier concentration around GBs has been attempted recently by F\^{a}te and Senatore by employing YBCO-based ionic-liquid gate transistors on 8$^\circ$ [001]-tilt SrTiO$_3$ bicrystal substrates\,\cite{Fate}. 
Such an electric double layer transistor is a powerful tool to inject a large carrier density of $\sim$8$\times$10$^{14}$\,cm$^{-2}$ at the material's surface\,\cite{Yuan}. The authors concluded that the level of carrier density by electrostatic doping was comparable to that of Ca-doping mentioned above. A similar attempt on tuning GB properties by an electric double layer transistor device has been reported by Hassan and Wimbush on La$_{1.85}$Sr$_{0.15}$CuO$_4$ by investigating the resistive transition\,\cite{Hassan}.

Another kind of GB engineering is the elimination of high-angle GBs. For this purpose, Jin $et$ $al$. invented a melt-process for YBCO wires\,\cite{Jin}. The resultant sample had a density of 6.2\,g/cm$^3$, which is 98\% of the theoretical density of YBCO. The self-field $J_{\rm c}$ of this melt-processed YBCO was almost doubled compared to sintered samples, however, $J_{\rm c}$ was decreased sharply by small magnetic fields of a few hundred gauss since high-angle GBs were not  removed completely.

To eliminate high-angle GBs or at least minimise their density in $RE$BCO bulk superconductors, the top-seeded-melt-growth process has been developed to grow quasi-single crystals\,\cite{Sawano}, which means that they contain secondary phase particles within the superconducting matrix and, therefore, are not true single crystals. $RE$BCO bulks can trap large magnetic flux densities, and this ability is determined by  $M=NJ_{\rm c}r$, where $M$ is the volume magnetization, $N$ is a geometrical constant, and $r$ is the diameter of the circulating supercurrent. Hence, a general processing aim in the fabrication of bulk $RE$BCO is the production of large-grain, weak-link-free samples with high $J_{\rm c}$. To date, the largest trapped field of 17.6\,T in bulk was recorded at 26\,K in a stack of two GdBCO bulks with 25\,mm in diameter\,\cite{Durrel-1} (Stacks of high-$T_{\rm c}$ tape showed a similarly high trapped field of 17.7\,T at 8\,K\,\cite{Anup}). Now several companies have commercialised $RE$BCO bulk superconductors. Applications by employing $RE$BCO bulk superconductors are well summarised in ref.\,\onlinecite{Durrel-2}.

Similar to $RE$BCO bulks fabricated by the top-seeded melt growth process, epitaxial thin films on single-crystal substrates do not contain high-angle grain boundaries. Additionally, growth-related defects such as screw dislocations\,\cite{Dam} and stacking faults that work as strong pinning centers result in high $J_{\rm c}$ values of more than 1\,MA/cm$^2$ at 77\,K. However, the question arose how to fabricate single-crystal-like $RE$BCO films in long lengths, since a powder-in-tube method is not applicable to fabricating $RE$BCO wires. This issue was solved by the development of technical substrates that have been used for the 2nd generation HTS wires, the $RE$BCO coated conductors. To date, mainly two kinds of technical substrates have been developed. One is the preparation of textured templates on Hastelloy C-276 tapes by ion beam assisted deposition (IBAD)\,\cite{Iijima}. IBAD yields biaxially textured buffer layers for epitaxial growth of $RE$BCO. The other is depositing epitaxial buffer layers on textured NiW tapes, which are thermo-mechanically treated (rolling-assisted biaxial texture, RABiTS)\,\cite{Goyal}. Thanks to the recent development of texture quality of those templates, the mean grain mis-orientation of $RE$BCO lies in the range of 3--5$^\circ$. Now tapes with high critical currents $I_{\rm c}$ and high uniformity of $\sim$1\,km length are being fabricated, and several companies have commercialised $RE$BCO coated conductors.
Table\,\ref{tab:GB_engineering} summarises GB engineering for $RE$BCO in various forms. As stated above, all methods are based on physical properties of $RE$BCO. Hence, it is important to reveal the intrinsic physical properties of superconductors to determine the strategy for tackling GB issues.

\begin{table}[t]
\caption{GB engineering for $RE$BCO in various forms.}
  \begin{tabular}{llr} \hline
Materials form      &	Methods                                                                  &	Refs. \\ \hline \hline
Bulks                    &  Employing melt-process, top-seeded-melt-growth  &  \onlinecite{Jin, Sawano} \\
Thin films             &  Ca-doping, electrostatic carrier doping  &  \onlinecite{Schmehl, Daniels, Hammerl, Fate} \\
Tapes                   &  Employing textured templates               &  \onlinecite{Iijima, Goyal} \\ \hline
\end{tabular}
\label{tab:GB_engineering}
\end{table}

\section{Grain boundary engineering of MgB$_2$}
\begin{figure}[b]
	\centering
		\includegraphics[width=7cm]{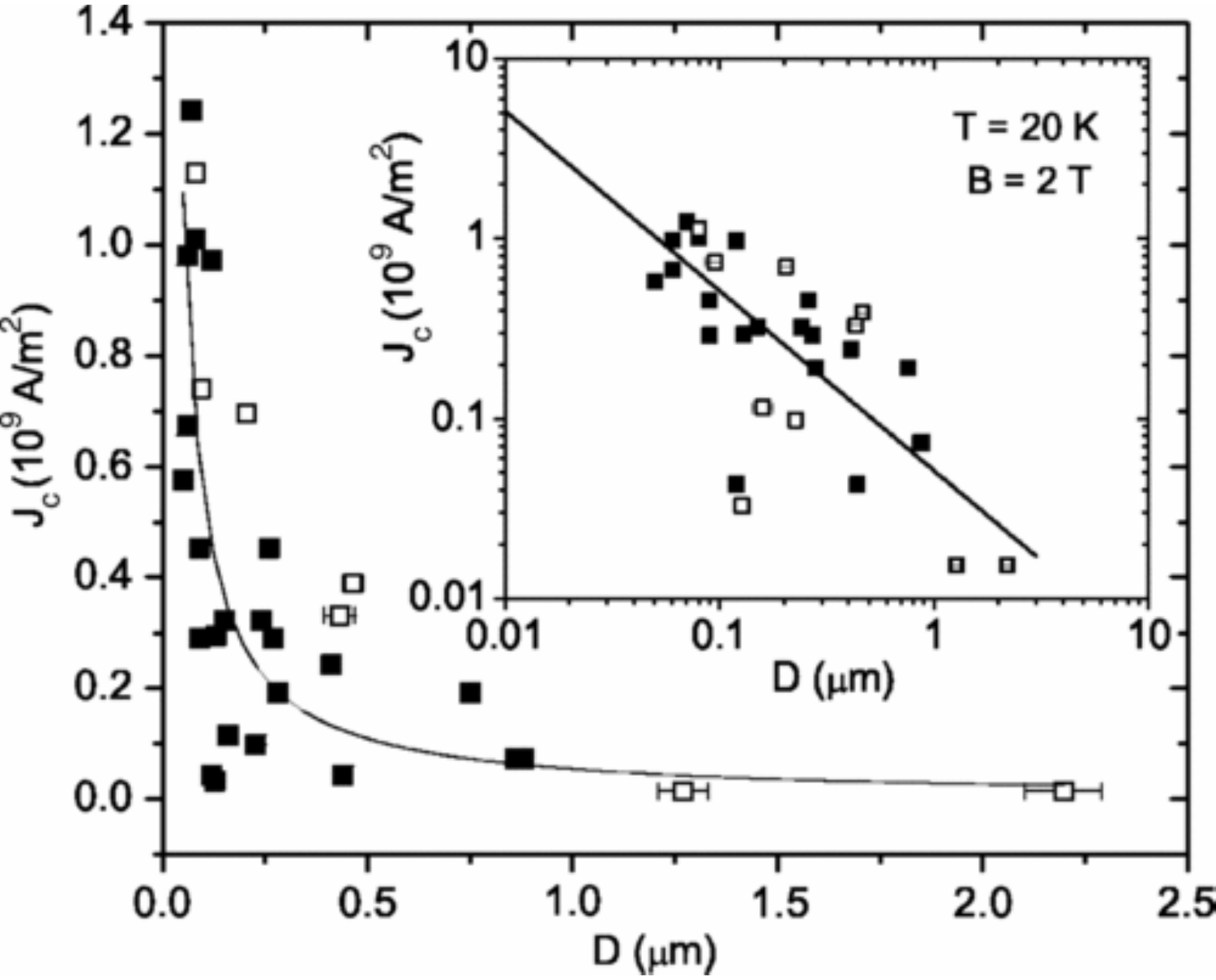}
		\caption{Grain size ($D$) dependence of $J_{\rm c}$ for dense MgB$_2$ samples. The line fits to the 1/$D$ dependence. The open symbols are non-doped samples and the solid symbols are doped samples. Reprinted with permission from\,\onlinecite{Martinez}. Copyright (2017) by the American Physical Society.} 
\label{fig:figure2}
\end{figure}

MgB$_2$ is a simple binary, metallic superconductor expected to be used at intermediate temperatures. Albeit MgB$_2$ has a rather high $T_{\rm c}$ of 39\,K and electromagnetic anisotropy of 2-5\,\cite{Buzea}, GBs do not show the weak-link limitation of transport current\,\cite{Larbalestier} due to the large coherence length and anisotropic $s$-wave order parameter symmetry. Soon after the discovery of superconductivity in MgB$_2$, high $J_{\rm c}$ was reported in untextured, polycrystalline wires produced by a powder-in-tube (PIT) process\,\cite{Dhalle, Glowacki, Grasso, Kumakura} owing to its “transparent” GBs. Like in other intermetallic compound superconductors, GBs belong to the predominant flux pinning centers in MgB$_2$\,\cite{Kitaguchi, Martinez, Collings, Matsushita, Senkowicz}. As shown in fig.\,\ref{fig:figure2}, a strong correlation between $J_{\rm c}$ and inverse grain size was observed\,\cite{Martinez}. Therefore, increasing the area density of GBs is an effective strategy to improve $J_{\rm c}$. On the other hand, GBs in MgB$_2$ can be easily covered with impurity phases, such as MgO and amorphous oxides. These insulating particles and/or wetting layers at GBs act as transport current limiters and reduce the connectivity, which deteriorates the inter-grain $J_{\rm c}$ significantly. In this section, recent activities on GB engineering of MgB$_2$ bulk samples, namely, controlling GB structure by doping/processing and enhancing GB density by grain size refinement and densification, are briefly introduced.

Refinement of MgB$_2$ grains has been extensively studied with various methods, including the use of fine boron powder\,\cite{Rosova}, mechanical refinement by milling\,\cite{Malagoli, Fujii-1, Fujii-2, Sugino}, low-temperature heat processing\,\cite{Yamamoto-2005, Kim-2007, Cai-2017, Cai-2018, Qaid}, and impurity doping with Ni\,\cite{Zhao-2017}, Cu\,\cite{Cheng-2017, Ma-2008, Ma-2009, Ma-2010}, Ag\,\cite{Shimoyama, Ozturk}, In\,\cite{Tachikawa}, and Hf\,\cite{Takahashi}. Ma $et$ $al$. reported that lamellar MgB$_2$ grains can be formed with Mg--Cu liquid, and these lamellar MgB$_2$ grains are much better connected than the MgB$_2$ grains of typical morphology\,\cite{Ma-2010}.

Another strategy for increasing the GB density per unit volume is densification, which also contributes to an improvement in connectivity by decreasing porosities. As densification techniques for bulk materials, high pressure synthesis and the magnesium diffusion method are commonly employed. To date, disk-shaped MgB$_2$ bulks prepared by a hot-pressing technique have been prepared by many groups\,\cite{Durrel-2012, Fuchs, Prikhna, Yamada}, and the resultant trapped field reached 3\,T\,\cite{Durrel-2012}. 
Nearly fully dense MgB$_2$ bulks can be obtained by spark plasma sintering in a relatively short time (less than an hour)\,\cite{Aldica, Badica-2016, Badica-2018, Berger, Murakami, Hassler}. The magnesium diffusion/infiltration technique is known as an alternative way to obtain dense MgB$_2$ wires/bulks without applying high pressure during heating\,\cite{Giunchi-2007, Giunchi-2011, Ueda-2005, Togano-2009, Li-2012}. 
Bhagurkar $et$ $al$. reported nearly fully dense undoped and carbon-doped MgB$_2$ bulks fabricated by the infiltration and growth process\,\cite{Bhagurkar-2016, Bhagurkar-2017, Bhagurkar-2018}. High trapped fields of $\sim$3\,T have been measured at 5\,K at the center of a stack of two bulk MgB$_2$ samples fabricated using this technique\,\cite{Bhagurkar-2016}. Owing to the progress in the technologies of producing relatively large, dense MgB$_2$ bulks, efforts in new bulk application fields, including magnetic shielding\,\cite{Gozzelino-2019}, magnetic levitation\,\cite{Savaskan-2019}, and biomedical applications\,\cite{Durrel-2}, as well as permanent-magnet applications have begun.

\section{$\theta_{\rm GB}$ dependence of inter-grain $J_{\rm c}$ for Fe-based superconducting bicrystal films} 
\subsection{NdFeAs(O,F)}
\begin{figure}[t]
	\centering
		\includegraphics[width=7cm]{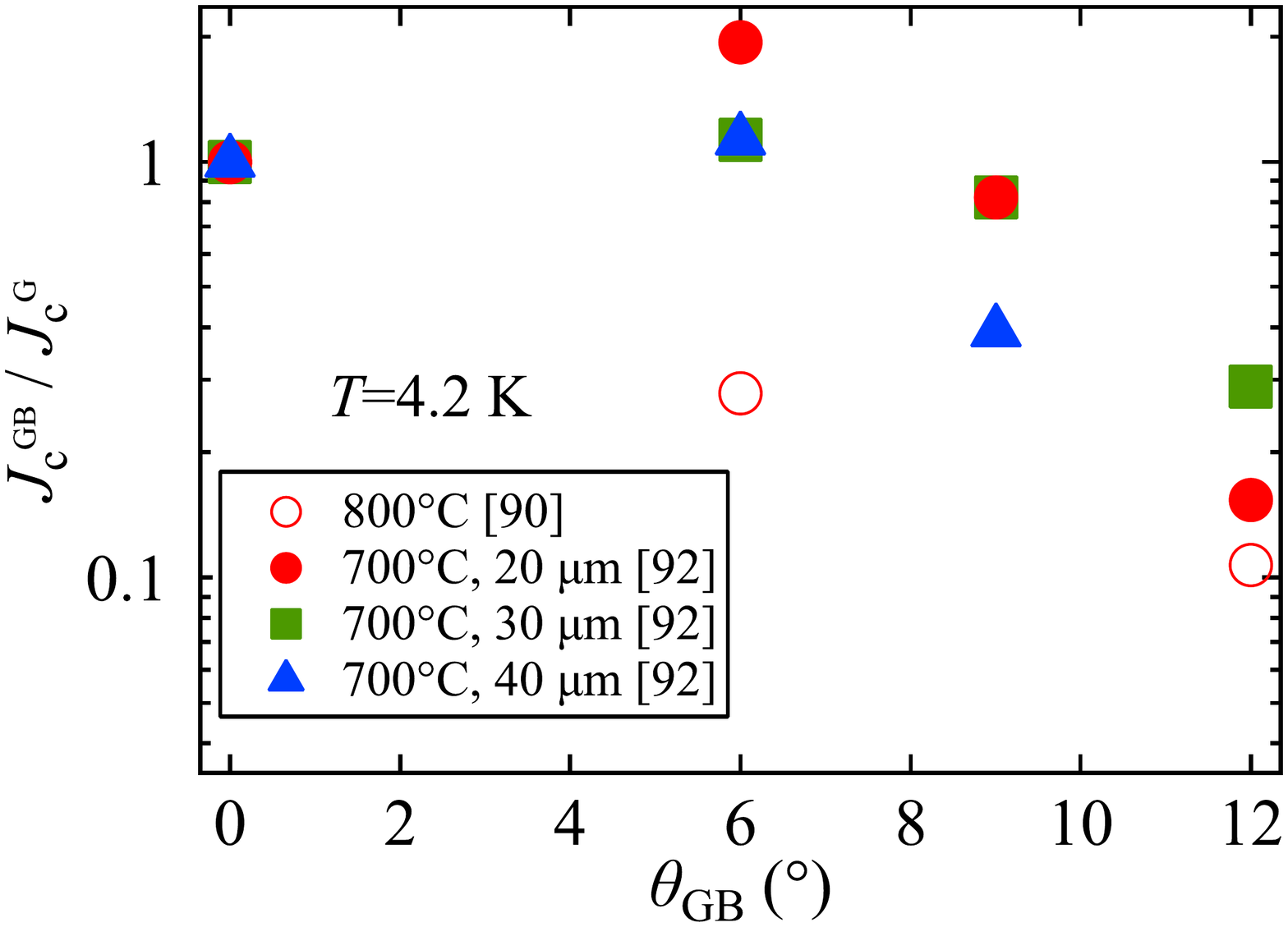}
		\caption{Angle dependence of the GB transparency $J_{\rm c}^{\rm GB}/J_{\rm c}^{\rm G}$ at 4.2\,K
		 for NdFeAs(O,F) bicrystal films with different deposition temperatures and thicknesses of the NdOF over-layer.} 
\label{fig:figure3}
\end{figure}

In the early stage of FBS research, LaFeAs(O,F)\,\cite{Aki-1, Haindl}, SmFeAsO$_{0.85}$, NdFeAs(O,F)\,\cite{Kametani}, and SmFeAs(O,F)\,\cite{Senatore} exhibited electromagnetic granularity similar to the cuprates. On the other hand, clear Meissner state due to bulk diamagnetism was observed in polycrystalline bulk $Ln$FeAs(O,F) ($Ln$: lanthanoide) samples by magneto-optical imaging at a magnetic field below the lower critical field $H_{\rm c1}$\,\cite{Aki-2}.
Moreover, the roof top pattern was observed in the remanent magnetic field distribution image of the field-cooled sample, suggesting the macroscopic critical state of the whole bulk. Although $J_{\rm c}$ was higher than that of the untextured cuprates, it was still one or two orders of magnitude lower than MgB$_2$, which does not show weak-link behaviour.
Detailed microstructural analysis revealed that the early polycrystalline samples were multiphase composites containing impurities and structural defects, such as cracks and an amorphous wetting phase\,\cite{Kametani}. GBs of these samples were random and mixed with tilt and twist grain boundaries. Additionally, clean grain boundaries (i.e. no impurities between the grains) were also confirmed. To understand the effect of intrinsic and extrinsic factors on GB transport properties, bicrystal experiments on $Ln$FeAs(O,F) with single GB have been long desired. Under such circumstances, Omura $et$ $al$. grew NdFeAs(O,F) films on MgO bicrystal substrates with [001]-tilt GB of various misorientation angles by molecular beam epitaxy\,\cite{Omura}. Here, NdFeAs(O,F) thin films have been fabricated via a two-step process: Parent NdFeAsO is fabricated and subsequently a NdOF overlayer is deposited on NdFeAsO for F-doping\,\cite{Kawaguchi-1}. Figure\,\ref{fig:figure3} shows the misorientation angle dependence of the GB transparency $J_{\rm c}^{\rm GB}/J_{\rm c}^{\rm G}$ ($J_{\rm c}^{\rm GB}$: inter-grain $J_{\rm c}$, $J_{\rm c}^{\rm G}$: intra-grain $J_{\rm c}$) for NdFeAs(O,F) fabricated on symmetric [001]-tilt MgO bicrystal substrates measured at 4.2\,K. The inter-grain $J_{\rm c}$  was reduced by 30\% compared to the intra-grain $J_{\rm c}$ even at $\theta_{\rm GB}$=6$^\circ$, indicating that the critical angle for those samples may be less than 6$^\circ$. However, the results show an extrinsic effect because NdFeAs(O,F) was damaged by excess F. The NdOF/NdFeAs(O,F) bilayer is similar to Y$_{0.7}$Ca$_{0.3}$Ba$_2$Cu$_3$O$_{7-\delta}$/YBCO. As known from the work of Y$_{0.7}$Ca$_{0.3}$Ba$_2$Cu$_3$O$_{7-\delta}$/YBCO bilayer\,\cite{Hammerl}, Ca preferentially diffused through the GB and enriched Ca around the YBCO grains. Similarly, F diffused along the GB in NdFeAs(O,F), however, the excess F deteriorated the NdFeAs(O,F) layers due to the strong reactivity.
To extract the intrinsic nature of the GBs in NdFeAs(O,F), the damage around the GBs had to be minimised. Hence, the NdOF overlayer was deposited at 750$^\circ{\rm C}$ instead of 800$^\circ{\rm C}$ in order to suppress the excess F diffusion along the GB.
With this method $J_{\rm c}^{\rm GB}/J_{\rm c}^{\rm G}$ is at nearly constant value of unity up to 8.5$^\circ$ (Figure\,\ref{fig:figure3})\,\cite{Iida-1}. Beyond 8.5$^\circ$, $J_{\rm c}^{\rm GB}/J_{\rm c}^{\rm G}$ decreases with $\theta_{\rm GB}$. Another distinct feature is that $J_{\rm c}^{\rm GB}/J_{\rm c}^{\rm G}$ is larger than 1 at $\theta_{\rm GB}$=6$^\circ$. A plausible reason for this phenomenon is GB pinning. This may explain the improvement of $J_{\rm c}$ for Co- and P-doped BaFe$_2$As$_2$ deposited on technical substrates\,\cite{Katase-1, Sato}, which is discussed later.

Lowering the deposition temperature for the NdOF overlayer certainly improved the GB properties of NdFeAs(O,F). However, both inter- and intra-grain $J_{\rm c}$ in self-field was of the order of 10$^5$\,A/cm$^2$\,\cite{Iida-1}, which is one order of magnitude lower than that of an ordinary NdFeAs(O,F) (i.e. deposition temperature of NdOF is 800$^\circ$C). Additionally, damage to the GB region by F is inevitable despite lowering the deposition temperature of NdOF. Therefore, studying F-free oxypnictide would be necessary. To date, polycrystalline $Ln$(Fe,Co)AsO\,\cite{Mendoza} and epitaxial SmFeAs(O,H)\,\cite{Matsumoto} thin films are available and, therefore, 
studying the GB angle dependence of $J_{\rm c}$ of bicrystal films of those compounds would be reasonable.

\subsection{Co- and P-doped BaFe$_2$As$_2$}
\begin{figure}[b]
	\centering
		\includegraphics[width=7cm]{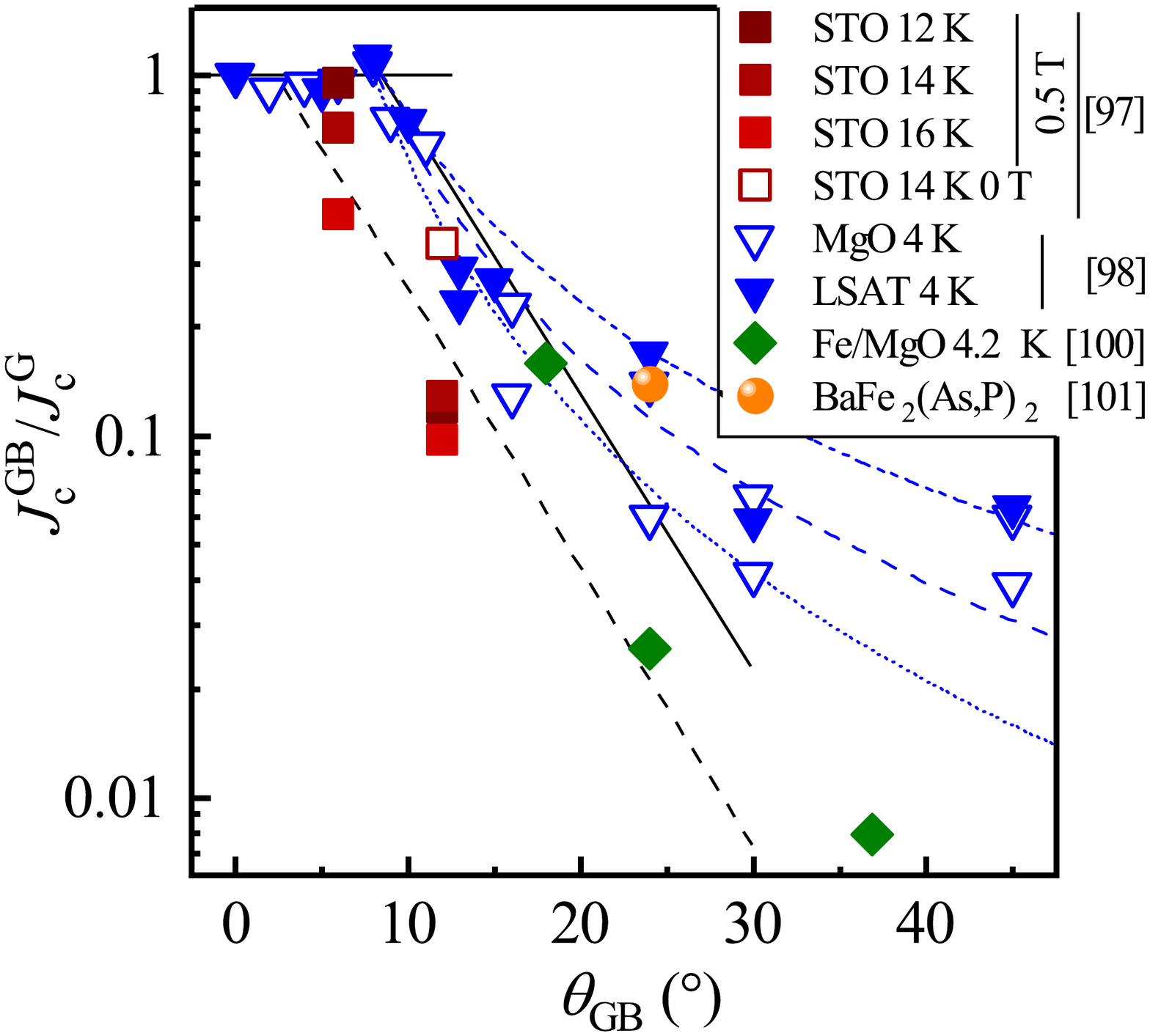}
		\caption{Angle dependence of the GB transparency for Ba(Fe$_{1-x}$Co$_x$)$_2$As$_2$ bicrystal films on different substrates [(LaAlO$_3$)$_{0.3}$(Sr$_2$TaAlO$_6$)$_{0.7}$ (LSAT), SrTiO$_3$ (STO) and Fe-buffered MgO (Fe/MgO)] grown by pulsed laser deposition and a BaFe$_2$(As$_{1-x}$P$_x$)$_2$ bicrystal film on MgO grown by molecular beam epitaxy. The lines are exponential (black) and power-law (blue) approximations of the data according to eqs.\,2 and 3, respectively.} 
\label{fig:figure4}
\end{figure}

Lee $et$ $al$. in 2009 were the first to show possible weak-link behaviour of [001]-tilt GBs in Ba(Fe$_{1-x}$Co$_x$)$_2$As$_2$ films ($T_{\rm c}$=20.5\,K) grown by pulsed laser deposition on bicrystalline SrTiO$_3$ substrates containing GBs with angles $\theta_{\rm GB}$ of 3$^\circ$, 6$^\circ$, 9$^\circ$, and 24$^\circ$\,\cite{Lee-1}. 
Also the low-angle GBs with $\theta_{\rm GB}$=6$^\circ$ and 9$^\circ$ showed weak-link behaviour in low-temperature laser scanning microscopy and magneto-optical imaging. This suggested a critical angle $\theta_{\rm c}< 6^\circ$, which later turned out to be in contrast to the results reported by Katase, Hiramatsu $et$ $al$.\,\cite{Katase-2, Hiramatsu} with a critical angle $\theta_{\rm c}\sim9^\circ$ from Ba(Fe$_{1-x}$Co$_x$)$_2$As$_2$ films on MgO ($T_{\rm c}$=20.7\,K) and LSAT ($T_{\rm c}$=21.6\,K) bicrystals. Presumably, an increased oxygen diffusion from SrTiO$_3$ under ultra high vacuum conditions along the GB leads to a slightly broader region of reduced order parameter around the GB which may have caused the $J_{\rm c}$ reduction of the low-angle GB\,\cite{Lee-1}. $T_{\rm c}$ was only minimally reduced across the 6$^\circ$ junction on SrTiO$_3$ by 0.2\,K with sharp resistive transition. Nevertheless, the elevated applied fields (500m\,T) and measurement temperatures (12, 14 and 16\,K) in\,\cite{Lee-1} should be regarded (data at low temperature and magnetic field are comparable to the ones reported in\,\cite{Katase-2, Hiramatsu}, fig.\,\ref{fig:figure4}). Later, Iida $et$ $al$. investigated high-angle GBs in Ba(Fe$_{1-x}$Co$_x$)$_2$As$_2$ films grown on Fe/spinel-buffered SrTiO$_3$ bicrystal substrates\,\cite{Iida-2}. The critical angle was around 8$^\circ$. The $J_{\rm c}$ reduction of the high-angle GBs, nevertheless, was 10 times larger than shown by Katase $et$ $al$, which again is most likely due to defects in the GB region because of the oxygen diffusion from SrTiO$_3$. Sakagami $et$ $al$. showed $J_{\rm c}^{\rm GB}$ as high as 1\,MA/cm$^2$ at 4\,K for $\theta_{\rm GB}$=24$^\circ$ for a BaFe$_2$(As$_{1-x}$P$_x$)$_2$ film on an MgO bicrystal grown by molecular beam epitaxy\,\cite{Sakagami}, which is due to the higher $T_{\rm c}$ and hence higher $J_{\rm c}^{\rm G}$ than for Co-doped BaFe$_2$As$_2$. Despite the low number of data points, $\theta_{\rm c}$ of BaFe$_2$(As$_{1-x}$P$_x$)$_2$  is most likely also larger than for the cuprates. The $J_{\rm c}(\theta_{\rm GB})$ dependence for GBs in Co- and P-doped BaFe$_2$As$_2$ as available in literature is summarised in Figure\,\ref{fig:figure4}. Clearly, $J_{\rm c}^{\rm GB}$ remains at $J_{\rm c}^{\rm G}$ up to $\theta_{\rm c}$$\sim$9$^\circ$. At higher angles, exponential decay of the GB transparency with a slope similar to that of the cuprates (dashed black line) is observed. $J_{\rm c}^{\rm GB}$ is, on the other hand, nearly constant for 30$^\circ$ and 45$^\circ$. An alternative possible phenomenological description of the data beyond $\theta_{\rm c}$ is shown as blue dashed lines according to

\begin{equation}
\frac{J_{\rm c}^{\rm GB}}{J_{\rm c}^{\rm G}} = \left(\frac{\theta_{\rm GB}}{\theta_0}\right)^{-n}
\end{equation}

\noindent
with sets of parameters ($\theta_0$, $n$) of (8.5, 1.7), (8.25, 2.05), and (8.0, 2.4) for high, medium and low $J_{\rm c}$ values, respectively. Such a function might be worthwhile for testing in statistical simulation models of GB networks such as in ref.\,\cite{Eisterer-1}, however the differences to the logarithmic fits may be small, especially for sharp textures.

Occasionally, pinning effects can be observed in single high-angle GBs, even though the absolute $J_{\rm c}$ value might not be higher than the intra-grain values. An example is found in ref.\,\cite{Lee-1} for the 24$^\circ$ GB junction of Ba(Fe$_{1-x}$Co$_x$)$_2$As$_2$ on SrTiO$_3$. 
Despite the lower $J_{\rm c}$ values for all applied fields, a peak effect is visible near 5\,T which can be attributed to pinning of vortices in the GB plane or surrounding region itself as well as by interaction with vortices in the grains nearby. In ref.\,\cite{Katase-2}, this effect can be recognised as well but is not discussed there. High-angle GBs can also supply pinning centres indirectly: Fe nanoparticles preferably grow in or near such GBs, see ref.\,\cite{Jens-2}, most likely because of higher diffusion rates in the grain boundaries or strain effects. $J_{\rm c}$ of the sample with these 45$^\circ$ GBs was increased in a wide range of field and temperature compared to a similar sample without 45$^\circ$ misorientations with a pronounced $c$-axis peak due to the $c$-axis-oriented GBs and the slightly $c$-axis-elongated Fe nanoparticles.

\subsection{Fe(Se,Te)}
\begin{figure}[t]
	\centering
		\includegraphics[width=7cm]{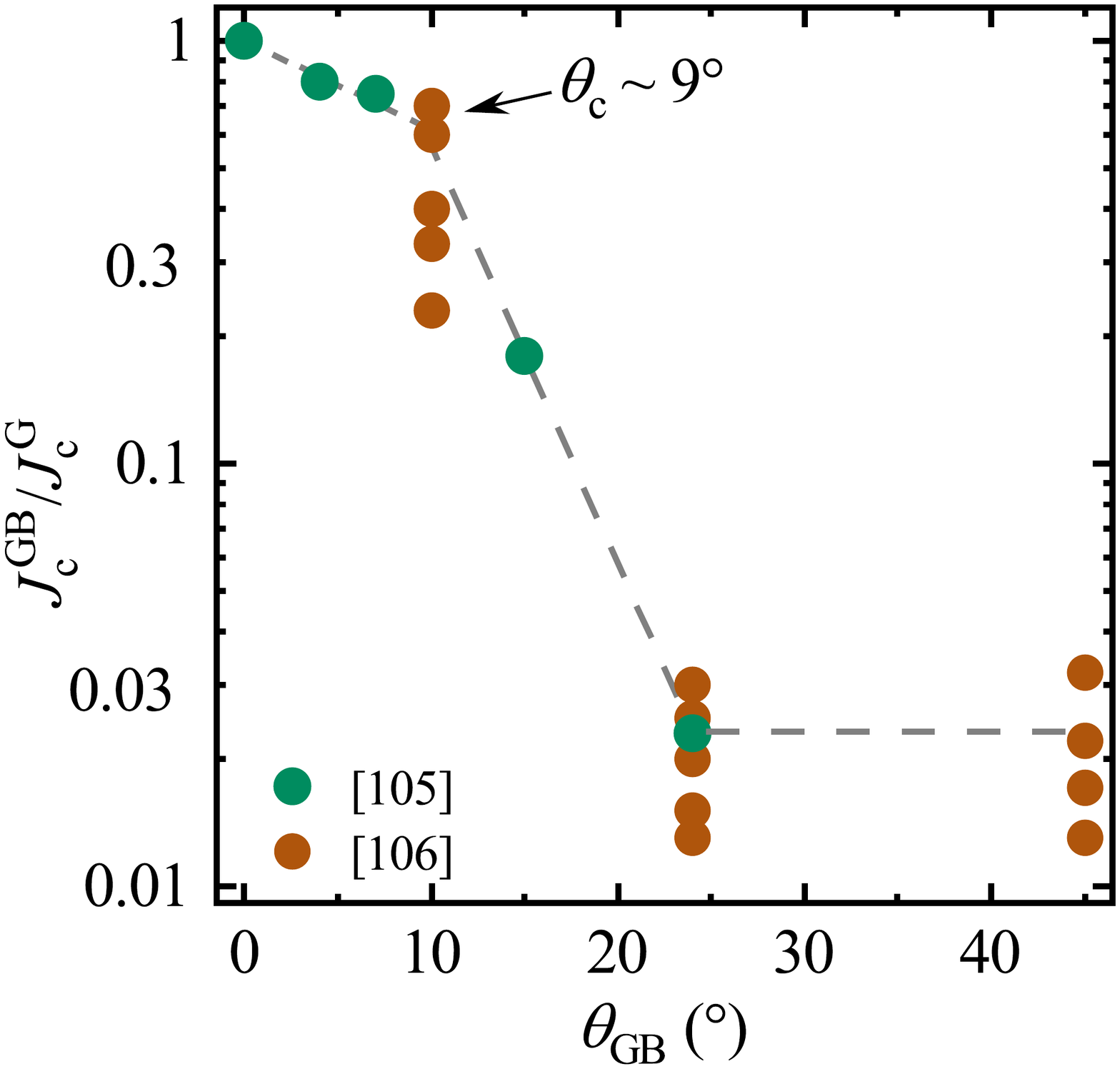}
		\caption{Angle dependence of the self-field GB transparency for Fe(Se,Te) bicrystal junctions at 4\,K.} 
\label{fig:figure5}
\end{figure}

A self-field inter-grain $J_{\rm c}$ of around 10$^4$\,A/cm$^2$ at low temperatures for Fe(Se,Te) of a 45$^\circ$ GB\,\cite{Sarnelli-1}, only 10 times lower than the intra-grain $J_{\rm c}$ (instead of 10$^4$ as for the cuprates) suggested a $J_{\rm c}$ decrease with $\theta_{\rm GB}$ not as strong as for the cuprates. $J_{\rm c}$ values being rather independent of the bridge width and the uniform Josephson current along the junction were attributed by Sarnelli $et$ $al$. to the $s$-wave symmetry in these materials. Later, the full $J_{\rm c}(\theta_{\rm GB})$ dependence with $\theta_{\rm c}\sim9^\circ$ was measured by Si $et$ $al$.\,\cite{Si-1} and Sarnelli $et$ $al$.\,\cite{Sarnelli-2}, see fig.\,\ref{fig:figure5}. As Si $et$ $al$. found, $J_{\rm c}(\theta_{\rm GB})$ maintains $\sim$10$^5$\,A/cm$^2$ at 4.2\,K for $\theta_{\rm GB}<7^\circ$ even in applied fields of around 10\,T, showing that these GBs are strong links. For $\theta_{\rm GB}$$>$15$^\circ$ the GBs showed significantly lower $J_{\rm c}$ values and a foot structure in the temperature dependence of the resistivity, $\rho(T)$, due to a larger distorted region near the GB. If intrinsic effects, such as thermally activated phase slippage\,\cite{Gross}, may contribute in such samples with low $T_{\rm c}$ values, is still to be investigated. In contrast to NdFeAs(O,F) and Ba(Fe$_{1-x}$Co$_x$)$_2$As$_2$, the GB transparency also decreased below $\theta_{\rm c}$, and no  increased values were found near $\theta_{\rm c}$. However, further experiments in the vicinity of the critical angle are necessary to clarify this. The level of $J_{\rm c}^{\rm GB}$ is almost constant above 24$^\circ$\,\cite{Sarnelli-2}.

\section{Grain boundary engineering of Fe-based superconductors}
\subsection{Grain boundary networks in Fe-based superconducting films}
Long-length YBCO coated conductors are fabricated on polycrystalline templates, which are in- and out-of-plane textured. Hence, thin films on technical substrates (e.g. Hastelloy) contain many low-angle GBs, which provides a good platform for studying how low-angle GB networks affect $J_{\rm c}$. To date, most of the FBS coated conductors as proof-of-principle studies have been fabricated on metallic templates with biaxially textured MgO prepared by IBAD. Fe(Se,Te) has been also deposited on CeO$_2$-buffered RABiTS. Here, we briefly review each FBS on technical substrates.

\subsubsection{$Ln$FeAs(O,F).}
So far, only NdFeAs(O,F) has been grown on a technical substrate among $Ln$FeAs(O,F). Iida, H{\"a}nisch $et$ $al$. reported on the fabrication of NdFeAs(O,F) on IBAD-MgO by molecular beam epitaxy\,\cite{Iida-3}. Structural characterization by X-ray diffraction revealed some 45$^\circ$ rotated grains, resulting in a low self-field $J_{\rm c}$ at 5\,K of 70\,kA/cm$^2$ (corresponding to a reduced temperature of $t$=$T/T_{\rm c,0}$=0.135), which is comparable to the 40\,kA/cm$^2$ at 4.2\,K measured for partially textured PIT SmFeAs(O,F) tape\,\cite{Zhang-1}. However, it was 1.5 times lower than $J_{\rm c}$ on MgO single crystal of 3\,MA/cm$^2$\,\cite{Chiara-1}. After optimisation of the deposition conditions for NdFeAs(O,F) on IBAD-MgO, rotated grains were successfully removed, resulting in a record self-field $J_{\rm c}$ for this kind of films of around 2\,MA/cm$^2$\,\cite{GuO}.

\subsubsection{Ba(Fe$_{1-x}$Co$_x$)$_2$As$_2$.}
Already in 2011, Ba(Fe$_{1-x}$Co$_x$)$_2$As$_2$ films were grown on Fe-buffered IBAD-MgO templates by pulsed laser deposition\,\cite{Iida-4}. The films were biaxially textured, and the crystalline quality of both Fe-buffer and superconducting layers with full width at half maximum values  of $\Delta \phi_{\rm Fe}$$\sim$5$^\circ$ and $\Delta \phi_{\rm Ba122}$$\sim$5$^\circ$ was comparable to that of the MgO ($\Delta \phi_{\rm MgO}$$\sim$6$^\circ$). The Ba(Fe$_{1-x}$Co$_x$)$_2$As$_2$ coated conductor showed an onset $T_{\rm c}$ of 22\,K 
($T_{\rm c,0}$=17.5\,K), which was a relatively large transition width. Due to the poor crystalline quality, the self-field $J_{\rm c}$ of 0.1\,MA/cm$^2$ at 8\,K ($t$=0.457) was 8-10 times lower than on single crystalline MgO. This electro-magnetic granularity in Ba(Fe$_{1-x}$Co$_x$)$_2$As$_2$ coated conductors was later suppressed by employing a sharply in-plane textured MgO ($\Delta \phi_{\rm MgO}$=2.4$^\circ$), leading to a $J_{\rm c}$ level comparable to the film on single crystalline substrate\,\cite{Trommler}. Shortly after, Katase $et$ $al$. grew Ba(Fe$_{1-x}$Co$_x$)$_2$As$_2$ on IBAD-MgO without a buffer layer\,\cite{Katase-1}. Thanks to the self-epitaxy effect, a sharp in-plane texture of $\Delta \phi_{\rm Ba122}$=3.2-3.5$^\circ$ was reached irrespective of the template's $\Delta \phi_{\rm MgO}$ values.

Although $J_{\rm c}$ is limited by GBs in general, GBs in FBS also have advantages: a) The larger $\theta_{\rm c}$ than for the cuprates (as reviewed above) allows for less-textured templates. b) GBs may contribute to flux pinning, resulting in high current carrying capabilities in magnetic field\,\cite{Katase-1}: Ba(Fe$_{1-x}$Co$_x$)$_2$As$_2$ films on IBAD-MgO of different in-plane full width at half maximum values ($\Delta \phi_{\rm MgO}$=5.5$^\circ$, 6.1$^\circ$, 7.3$^\circ$) lead to Ba(Fe$_{1-x}$Co$_x$)$_2$As$_2$ texture spreads of $\Delta \phi_{\rm Ba122}$=3.1$^\circ$, 3.2$^\circ$, and 3.5$^\circ$. Apparently, the self-field $J_{\rm c}$ (1.2, 1.6, 3.6\,MA/cm$^2$) is correlated to $\Delta \phi_{\rm Ba122}$, which is due to $\theta_{\rm GB}$ of most GBs being below $\theta_{\rm c}$ (i.e. they are strong links) and due to an increasing dislocation density with misorientation spread. One of the films on IBAD-MgO ($\Delta \phi_{\rm MgO}$=6.1$^\circ$) showing higher $J_{\rm c}$($B$$\parallel$$c$) values than a film on MgO single crystal for all fields at
low and medium temperature and even higher than $J_{\rm c}$($B$$\parallel$$ab$) up to a 
temperature-dependent cross-over field supported this finding.

Xu $et$ $al$. reported recently on the growth of Ba(Fe$_{1-x}$Co$_x$)$_2$As$_2$ on SrTiO$_3$/LaMnO$_3$-buffered IBAD-MgO\,\cite{Xu-1}. Almost isotropic $J_{\rm c}$ of 0.86 and 0.96\,MA/cm$^2$ at 9\,T and for $B$$\parallel$$c$ and $B$$\parallel$$ab$, respectively, were recorded at 4.2\,K.

To understand the $J_{\rm c}$-$B$ properties of FBS grown on technical substrates, Abrikosov-Josephson vortices should also be taken into account. Abrikosov vortices with Josephson cores, so-called Abrikosov-Josephson vortices, are present in low-angle GBs for YBCO\,\cite{Gurevich-2, Horide}. Although there are no reports on Abrikosov-Josephson vortices in FBS, they might be present in GB networks having small misorientation angles. As has been reported by Palau $et$ $al$. for cuprates\,\cite{Palau}, the inter-grain $J_{\rm c}$ correlates with intra-grain $J_{\rm c}$ due to the magnetic interaction between Abrikosov-Josephson vortices at low-angle GBs and Abrikosov vortices in the grains. Hence, it is possible to improve the pinning potential of Abrikosov-Josephson vortices by increasing the density of strong pinning centres in close vicinity of the GB region.

\subsubsection{BaFe$_2$(As$_{1-x}$P$_x$)$_2$.}
\begin{figure}[b]
	\centering
		\includegraphics[width=12cm]{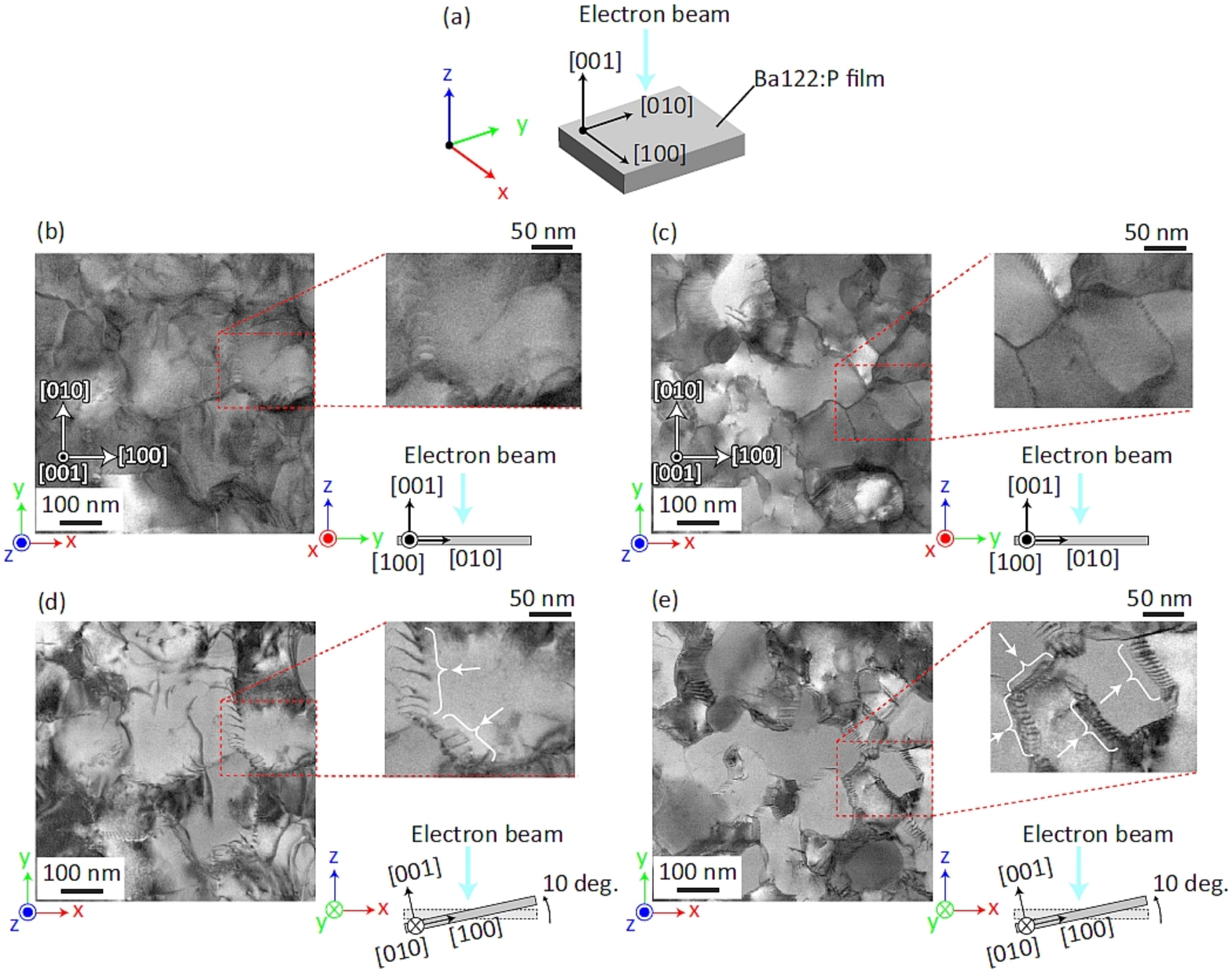}
		\caption{Plane-view bright-field scanning TEM images of BaFe$_2$(As$_{1-x}$P$_x$)$_2$ films on two IBAD templates with $\Delta \phi_{\rm MgO}$=4$^\circ$ and 8$^\circ$. 
		(a) Relationship between film orientation, incident electron beam, and global axes (X, Y, and Z). (b, c) Typical plane-view images by normal electron beam incidence 
		for $\Delta \phi_{\rm MgO}$=4$^\circ$ (b) and 8$^\circ$ (c). (d, e) Slanted-angle images (tilting by 10$^\circ$) for visualizing dislocations at same areas of 
		(b, c), respectively. The arrows in the top right image of (d, e) show arrays of dislocations. Reproduced with permission from Sato $et$ $al$., $Sci$. $Rep$. {\bf6},
		 36828 (2016). Copyright 2016 Spring Nature.} 
\label{fig:figure6}
\end{figure}

Sato $et$ $al$. grew BaFe$_2$(As$_{1-x}$P$_x$)$_2$ on IBAD-MgO with $\Delta \phi_{\rm MgO}$=4$^\circ$ and 8$^\circ$\,\cite{Sato} by pulsed laser deposition. Energy dispersive X-ray spectroscopy in TEM cross sectional images showed clean and homogeneous GBs. The grain size of $\sim$100\,nm as well as dislocation arrays in the GB regions were seen in plane-view TEM images (fig.\,\ref{fig:figure6}). The crystalline quality of the film on the 4$^\circ$ substrate was better than that of the film on MgO with $\Delta \phi_{\rm MgO}$=8$^\circ$, and its $T_{\rm c,0}$ of 23\,K was 4\,K higher than for the 8$^\circ$ substrate. Nevertheless, the film on the 8$^\circ$ substrate showed a larger $c$-axis peak and 1.5 times higher $J_{\rm c}$ values than the 4$^\circ$ template, which is due to the larger density of especially vertical defects related to the GB networks.
Since it has a higher $J_{\rm c}$ and therefore also irreversibility line, the 4$^\circ$ sample surpasses the 8$^\circ$ sample at a temperature-dependent crossover field. However, the 8$^\circ$
sample still shows the stronger $c$-axis peak. One of these samples on 8$^\circ$ template was further investigated in high fields\,\cite{Iida-5}. The field-dependent pinning force densities, $F_{\rm p}(H)$, analysed with the modified Dew-Hughes model, suggest that pinning is strongly dominated by surface pinning, i.e. on 2-dimensional defects in the matrix such as dislocation networks and full GBs. The $V(I)$ curves clearly showed a crossover between GB limitation (non-ohmic linear differential) at low fields and and pinning limitation (power law behaviour, representing flux creep effect) at high fields. Similar non-ohmic linear differential behaviour has been measured for Ba(Fe$_{1-x}$Co$_x$)$_2$As$_2$ with 45$^\circ$ and (110) misoriented grains\,\cite{Rodriguez} and polycrystalline SmFeAsO$_{0.85}$\,\cite{Kametani-1}.

A first 10\,cm long BaFe$_2$(As$_{1-x}$P$_x$)$_2$ tape was fabricated by pulsed laser deposition with a reel-to-reel system at a tape travelling speed of 
6\, mm/min\,\cite{Hosono-1}, and its  $T_{\rm c,0}$ was reduced $\sim$17\,K, which is slightly low compared to the static samples ($T_{\rm c,0}$$\sim$20\,K). $I_{\rm c}$ was 0.47\,mA/cm-width at 4.2\,K, corresponding to $J_{\rm c}$=47\,kA/cm$^2$. For enhancing $T_{\rm c}$ and $J_{\rm c}$, Fe$_3$P/BaFe$_2$(As$_{1-x}$P$_x$)$_2$ bilayers were fabricated on IBAD-MgO. This led to improved $T_{\rm c,0}$ of 24\,K and $I_{\rm c}$ of 975\,mA/cm-width ($J_{\rm c}$=175\,kA/cm$^2$) at 4.2\,K.

\subsubsection{Fe(Se,Te).}
Si $et$ $al$. were able to deposit high quality Fe(Se,Te) films on IBAD-MgO despite the relatively large mismatch of 9.5\%\,\cite{Si-2}. The in- and out-of-plane full width at half maximum values of Fe(Se,Te) were $\Delta \phi_{\rm Fe(Se,Te)}$=4.5$^\circ$
($<$$\theta_{\rm c}$) and $\Delta \omega_{\rm Fe(Se,Te)}$= 3.5$^\circ$. Since also [010]-tilt GBs in FBS probably have a similar $\theta_{\rm c}$ value, the GBs do not act as weak-links in these films. The lower $T_{\rm c,0}$ of 11\,K compared to single crystalline LaAlO$_3$ substrate ($\sim$15\,K) can be attributed to the large misfit and reduces self-field $J_{\rm c}$ at 4\,K to 0.2\,MA/cm$^2$. Xu $et$ $al$. grew Fe(Se,Te) films on LaMnO$_3$/IBAD-MgO \,\cite{Xu-2}.
The $T_{\rm c}$ of 15.8\,K was slightly enhanced due to the in-plane lattice compression. Whereas $\Delta \omega_{\rm Fe(Se,Te)}$=3.4$^\circ$ was similar to the film in ref.\,\cite{Si-2}, $\Delta \phi_{\rm Fe(Se,Te)}$=7.8$^\circ<\theta_{\rm c}$ was larger. The self-field $J_{\rm c}$ at 4.2\,K was 0.43\,MA/cm$^2$ nonetheless.

Not only IBAD-MgO templates but also CeO$_2$-buffered RABiTS were used for the growth of Fe(Se,Te) films\,\cite{Si-3}. The relevant lattice parameter of CeO$_2$ is around 3.82\,${\rm \AA}$, hence close to the  in-plane parameters of Fe(Se,Te). Si $et$ $al$. achieved films on RABiTS with a high $T_{\rm c}$ of $\sim$20\,K and sharp resistive transition, which may be due to the small lattice mismatch and absence of intercalated Fe. Even though the in-plane full width at half maximum value was $\sim6^\circ$, (corresponding to the underlying template of $\Delta \phi_{\rm CeO2}$=7$^\circ$), a large self-field $J_{\rm c}$ of 1.5\,MA/cm$^2$ at 4.2\,K was measured. According to the authors, the CeO$_2$ buffer layer is more important for the superconducting properties than the texture spread in this system. That means again GBs with angles up to 7$^\circ$ do not obstruct the current flow. To reduce the necessary number of buffer layers, Sylva $et$ $al$. recently deposited Fe(Se,Te) on Ni5W with a single CeO$_2$ buffer layer by pulsed laser deposition\,\cite{Sylva}. This sample showed $T_{\rm c}\sim$18\,K  (due to slight Ni poisoning) and a self-field $J_{\rm c}\sim$0.1\,MA/cm$^2$ as well as $J_{\rm c}(H)>20\,{\rm kA/cm}^2$ up to 18\,T at 4.2\,K.

\subsection{$Ae$Fe$_2$As$_2$ ($Ae$=Sr and Ba) polycrystalline samples}
The $Ae$-122 system is one of the most actively researched FBS for magnet applications owing to its moderately high $T_{\rm c}$ (38\,K for K-doped Ba-122), high $B_{\rm c2}$ (above 50\,T) and small electromagnetic anisotropy\,\cite{Hosono, Yao}. In recent years, high-$J_{\rm c}$ wires\,\cite{Weiss-1, Zhang-2, Gao}, pancake coil tests\,\cite{Wang}, and demonstration of bulk magnet trapping over 1 T\,\cite{Weiss-2,Iida-6, Ainslie} have been reported. The synthesis and GB engineering guidelines for the $Ae$-122 system fall into two different categories: uniaxial texturing of polycrystals, and making randomly oriented polycrystals without texturing. The former and latter guidelines are similar to those for Bi$_2$Sr$_2$Ca$_2$Cu$_3$O$_{10+\delta}$ tapes and MgB$_2$, respectively. For tape wires with their flat geometry produced by the PIT method, a uniaxially aligned microstructure can be obtained relatively easily by employing plate-like $Ae$-122 powder and cold working. On the other hand, for untextured polycrystalline forms such as bulks or round wires, high transport $J_{\rm c}$ can be successfully achieved by grain boundary engineering, which is discussed below.

\begin{figure}[b]
	\centering
		\includegraphics[width=7cm]{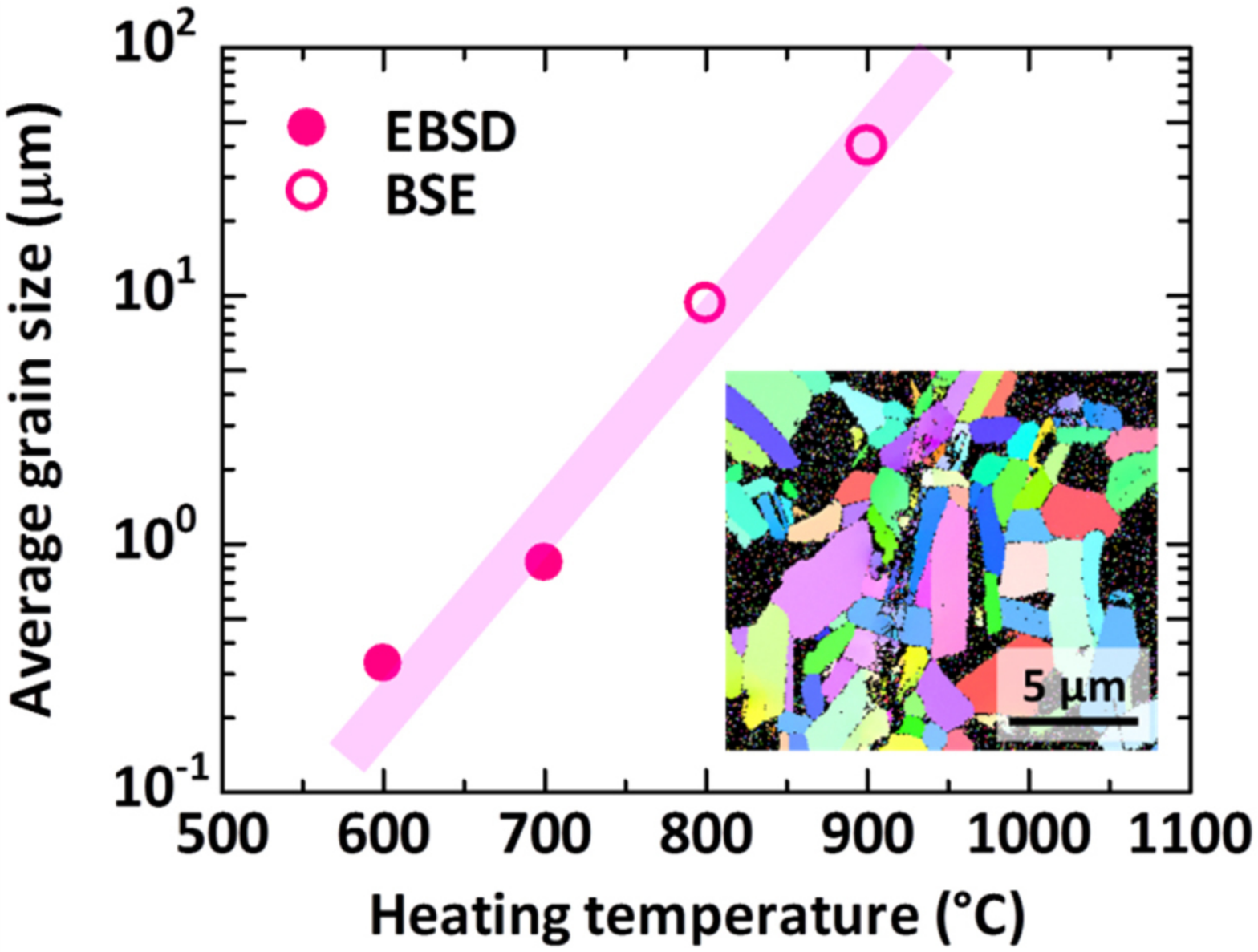}
		\caption{Clear correlation between average grain size and processing temperature can be seen for polycrystalline Co-doped Ba-122. The inset displays the inverse pole figure map obtained by scanning electron microscope EBSD for a sample heated at 700$^\circ$C for 48\,h. BSE stands for back-scattered electron. Reproduced from\,\onlinecite{Shimada}.} 
\label{fig:figure7}
\end{figure}

High transport $J_{\rm c}$ in Ag-clad Sr$_{0.6}$K$_{0.4}$Fe$_2$As$_2$ tapes prepared by the $ex$-$situ$ PIT method was reported by Zhang $et$ $al$\,\cite{Zhang-2}. To produce Sr-122 powder, Sr fillings, K pieces, and Fe and As powders were mixed and heat-treated at 900$^\circ$C for 35\,h. In order to improve the grain connectivity, 5wt\% Sn was added. After packing the powder into Ag tubes, drawing, flat rolling and hot pressing at 850$^\circ$C under 30\,MPa, Sr-122 tapes with 0.4\,mm in thickness were obtained. $J_{\rm c}$ of the Sr-122 tape was over 10$^5$\,A/cm$^2$ in 10\,T at 4.2\,K, and still remained at 8.4$\times$10$^4$\,A/cm$^2$ up to 14\,T. The high $J_{\rm c}$ in the Sr-122 tapes may be attributed to the combination of improved grain connectivity, grain texture, and inherently strong pinning. The $c$-axis orientation factor for the tape was 0.52, suggesting grain texturing after pressing. Such grain texturing is considered to alleviate weak-link behaviour at GBs and can also be achieved by high pressure sintering\,\cite{Pyon-1, Pyon-2, Tamegai} and conventional cold mechanical deformation processes\,\cite{Yao, Gao, Togano-1, Huang, Liu, Xu-3}. Ag-based Sn binary alloys were deployed for $ex$-$situ$ Ba-122 tapes as sheath material by Togano $et$ $al$\,\cite{Togano-2}. The Ag-Sn alloy has higher mechanical strength compared to pure Ag. Successful densification and texturing of the Ba-122 core were obtained. The Ag-Sn-sheathed Ba-122 tapes showed higher transport $J_{\rm c}$ compared to those of Ag-sheathed tapes. The authors further pointed out the significantly improved smoothness of the interface between the sheath and Ba-122 core. These findings indicate the benefit of using high mechanical strength sheath. 
Surprisingly high $J_{\rm c}$ values in randomly oriented, fine-grain wires and bulks were reported by Weiss $et$ $al$\,\cite{Weiss-1}. Sub-micron size Ba-122 powders were prepared by a mechanochemical reaction\,\cite{Weiss-3}. TEM showed that the average grain size is approximately 200\,nm. The randomly oriented polycrystalline structure contained many high-angle grain boundaries including clean and well connected ones. $J_{\rm c}$ at 4\,K reached 1.2$\times$10$^5$\,A/cm$^2$ and 2$\times$10$^4$\,A/cm$^2$ for Ba-122 wires with K-doping and Co-doping, respectively. Quite recently, the additional effect by ball milling was reported by Tokuta and Yamamoto\,\cite{Tokuta}: high-energy ball milling introduces lattice defects in Co-doped Ba-122, resulting in high $H_{\rm c2}$.

\begin{figure}[t]
	\centering
		\includegraphics[width=7cm]{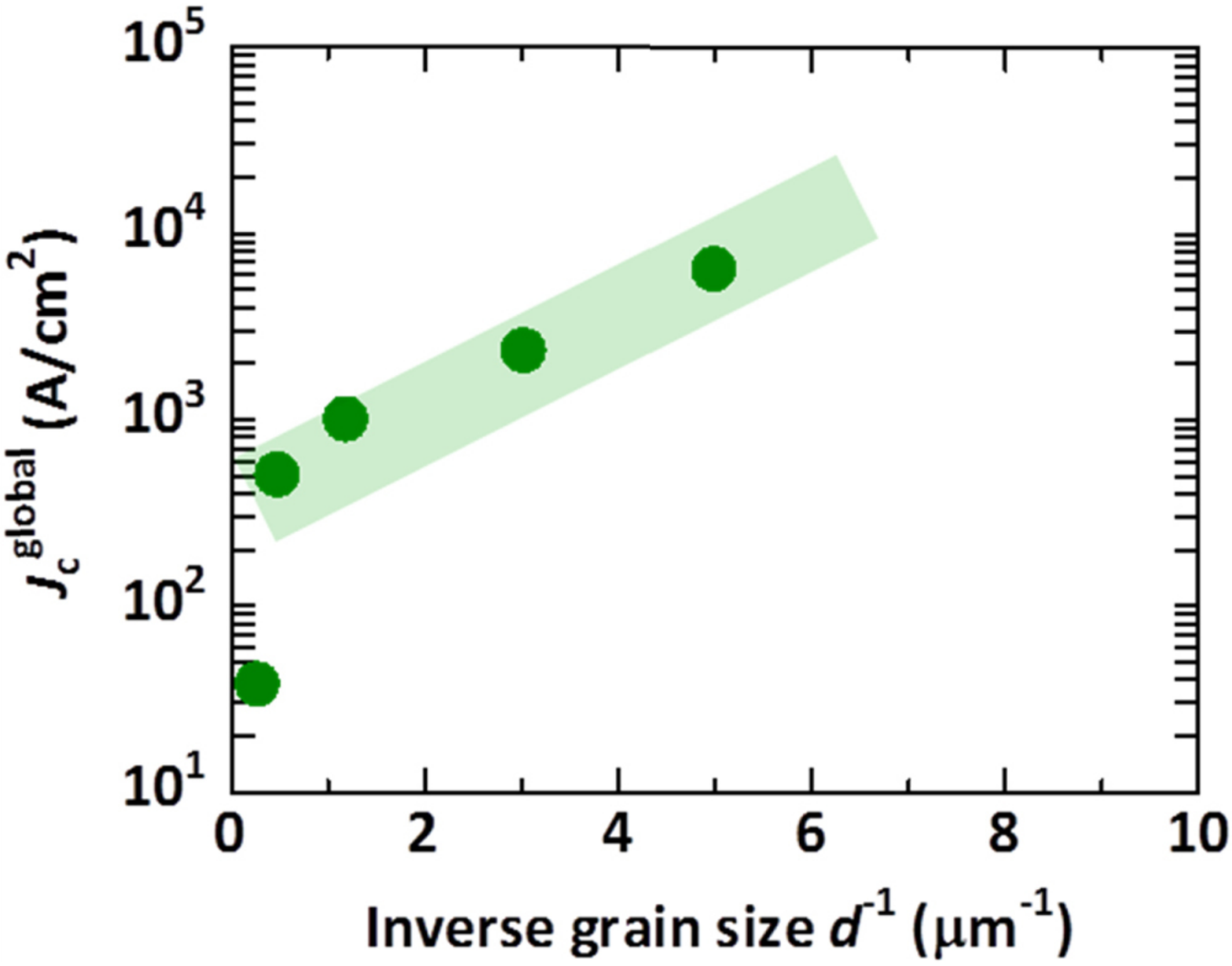}
		\caption{Inverse grain size (1/$d$) dependence of the global $J_{\rm c}$ for Co-doped Ba-122 polycrystalline samples. Reproduced from\,\onlinecite{Shimada}.} 
\label{fig:figure8}
\end{figure}

The influence of grain size, especially the benefit of small grains, was studied by Hecher $et$ $al$. based on the Josephson junction network model\,\cite{Hecher}. Scanning Hall probe microscopy and magnetic hysteresis measurements on Ba-122 bulks with systematically varied grain size showed that the field dependence of untextured polycrystalline samples can be significantly reduced when the grain size is small. Shimada $et$ $al$. reported on detailed microstructural investigations of the  Ba-122 bulk samples prepared at different heating temperatures\,\cite{Shimada}. A strong heating temperature dependence of the average grain size was observed for the temperature range between 600$^\circ$C and 900$^\circ$C (fig.\,\ref{fig:figure7}). Interestingly, a relationship between inverse grain size and magnetization $J_{\rm c}$ was confirmed, as shown in fig.\,\ref{fig:figure8}, which also supports that a small grain size would be preferable for randomly oriented polycrystalline Ba-122 samples.

Another key issue to address is chemical composition at grain boundaries. Kim $et$ $al$. reported grain and grain boundary composition in K-doped Ba-122 polycrystals by atom-probe tomography\,\cite{Kim}. Significant compositional variation and segregation of oxygen impurities at grain boundaries were observed. These would suppress transport current across grain boundaries. Detailed microstructural and electromagnetic investigation on artificial grain boundaries in bicrystal and natural grain boundaries in polycrystalline samples are very helpful for grain boundary engineering in this system.

\section{Discussion and Summary} 
Artificial, well-defined single grain boundaries have been 
used for fundamental investigations such as $J_{\rm c}$ transparency as a function of mis-orientation angle $\theta_{\rm GB}$.
For [001]-tilt symmetric bicrystal films, Fe(Se,Te), Co-doped Ba-122 and NdFeAs(O,F) showed the same critical angle $\theta_{\rm c}$ of around 9$^\circ$, which is larger than for cuprates. The dominant factor for governing $\theta_{\rm c}$ may be the symmetry of the order parameter. The reduction of current-carrying cross-section of GB regions by strain effects also controls the $\theta_{\rm c}$ as discussed by Gurevich and Pashitskii.\,\cite{Gurevich-1}, since the sensitivity of superconductivity by strain is different for FBS and cuprates. Additionally, the local off-stoichiometry near GBs affects the size of the normal core regions\,\cite{Gurevich-1}.
Although $\theta_{\rm c}$ is different for FBS and cuprates, both superconductors showed an exponential decay of inter-grain $J_{\rm c}$ with nearly the same slope for medium angles. However, for $\theta_{\rm GB}>24^\circ$ the inter-grain $J_{\rm c}$ for Co-doped Ba-122 and Fe(Se,Te) is almost constant, whereas such behaviour is absent for cuprates. Unlike YBCO, the microscopic understanding of GBs is still limited. To understand the GB characteristics, therefore, detailed microstructural analyses should be carried out to map out the local strain around the GBs. To rule out local off-stoichiometry,  bicrystal experiments using stoichiometric superconductors (e.g. LiFeAs and FeSe  or even CaKFe$_4$As$_4$) would be interesting for further studies.

\begin{figure}[b]
	\centering
		\includegraphics[width=7cm]{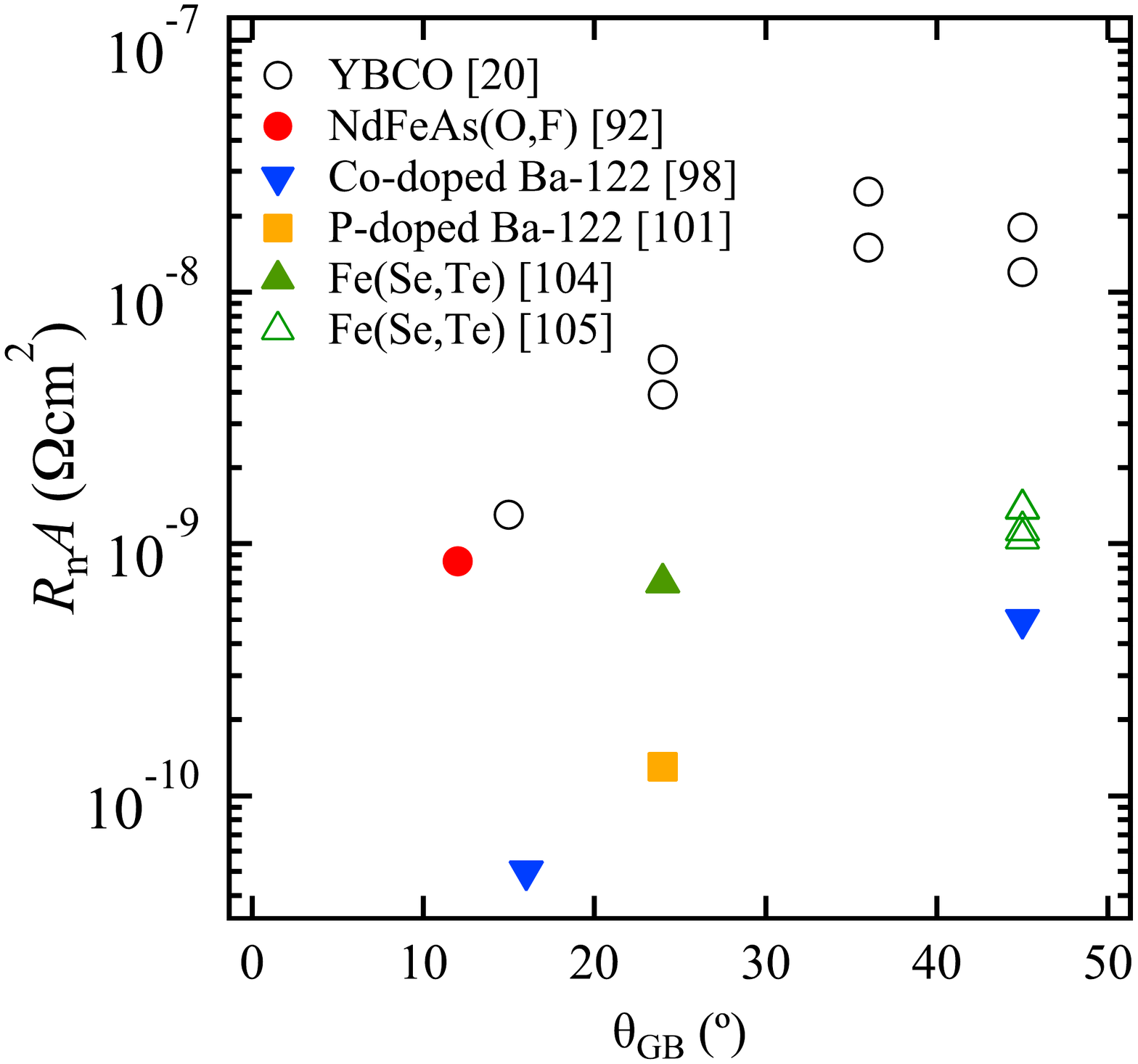}
		\caption{Comparative $\theta_{\rm GB}$ dependence of resistance area product $R_{\rm n}A$ at low temperature for YBCO at 4.2\,K\,\cite{Hilgenkamp-2}, 
		NdFeAs(O,F) at 4.2\,K\,\cite{Iida-1}, Ba(Fe$_{1-x}$Co$_x$)$_2$As$_2$ (Co-doped Ba-122 at 4\,K\,\cite{Katase-2}), BaFe$_2$(As$_{1-x}$P$_x$)$_2$
		 (P-doped Ba-122 at 2\,K\,\cite{Sakagami}), and Fe(Se,Te) at 4.2\,K\,\cite{Sarnelli-1, Si-1}.} 
\label{fig:figure9}
\end{figure}

Another distinct feature is that the product of the grain-boundary normal state resistance $R_{\rm n}$ and the grain boundary area $A$ for FBS is one or two orders of magnitude lower than that for cuprates due to the metallic nature of GBs (fig.\,\ref{fig:figure9}). Here, $R_{\rm n}$ corresponds to the slope of $V$-$I$ in the non-ohmic linear differential region and $A$ is the cross sectional area of the microbridge for transport measurements. For YBCO, band bending creates charge depletion layers around the GBs, which leads to the formation of insulating layers (Mott insulator)\,\cite{Mannhart}. The presence of insulating layers at large-angle GBs was experimentally confirmed by Winkler $et$ $al$. on YBCO bicrystal Josephson junctions\,\cite{Winkler}. For doped FBS, even if the depletion layers are formed at GBs by some mechanisms like strain effects, GB regions stay in the metallic state. This may be a part of the reason why the GBs of FBS are of metallic nature.

The critical angle as well as the $\theta_{\rm GB}$ dependence of the inter-grain $J_{\rm c}$ could be different for other types of bicrystals (i.e. [010]-tilt roof and valley, and twist). Hence, various types of bicrystal experiments should be carried out in order to deeply understand the characteristics of GBs, which is very important for the applications using a polycrystalline form such as PIT wires and bulk magnets.

Regarding applications, bicrystal experiments on K-doped Ba-122, which is believed to be an important material, should be carried out.
Experiments on single GBs in polycrystalline K-doped Ba-122 films have been reported\,\cite{Hong}. However, the information on well-defined, single GBs is necessary for deep understanding of GB characteristics. Due to the difficulty of the thin film growth, no studies on epitaxial K-doped Ba-122 thin films have been reported to date. Very recently, we have successfully fabricated epitaxial K-doped Ba-122\,\cite{Tony}, which opens an avenue for such bicrystal experiments.

Clean GBs without wetting phase may not become obstacles for inter-grain super-currents irrespective of the mis-orientation angle, which is similar to MgB$_2$. Even a high GB density acts as pinning centre ensemble.

From the above, grain boundary properties of FBS may be also located in between MgB$_2$ and cuprates.
Nevertheless, fundamental understanding of GB properties leads to further improvement of the current carrying capability of FBS by GB engineering.    
  
\begin{acknowledgments}
This work was supported by the JSPS Grant-in-Aid for Scientific Research (B) Grant Number 16H04646 as well as JST CREST Grant Number JPMJCR18J4. 
\end{acknowledgments}

\end{document}